\documentclass[12pt]{article}
\usepackage{amsmath,amssymb}
\usepackage{graphicx}
\newcommand{\eqdef}{\stackrel{\text{def}}{=}}

\newcommand{\sech}{\text{sech}}
\newcommand{\n}{\nonumber\\}

\newcommand{\ignore}[1]{}
\numberwithin{equation}{section}
\newcommand{\Romannumeral}[1]{\uppercase\expandafter{\romannumeral#1}}

\allowdisplaybreaks[4]

\begin{document}

\baselineskip=20pt

\newcommand{\preprint}{
\vspace*{-20mm}
}
\newcommand{\Title}[1]{{\baselineskip=26pt
   \begin{center} \Large \bf #1 \\ \ \\ \end{center}}}
\newcommand{\Author}{\begin{center}
   \large \bf  Ryu Sasaki \end{center}}
\newcommand{\Address}{\begin{center}
      Department of Physics, Shinshu University,\\
     Matsumoto 390-8621, Japan
    \end{center}}
\newcommand{\Accepted}[1]{\begin{center}
   {\large \sf #1}\\ \vspace{1mm}{\small \sf Accepted for Publication}
   \end{center}}

\preprint
\thispagestyle{empty}

\Title{Exactly solvable potentials  with finitely many discrete eigenvalues   
                  of arbitrary choice }                     %

\Author

\Address
\vspace{1cm}

\begin{abstract}
We address the problem of possible deformations of exactly solvable potentials 
having finitely many discrete eigenvalues of arbitrary choice.
As Kay and Moses showed in 1956, reflectionless potentials in 
one dimensional quantum mechanics are exactly solvable.
With an additional time dependence these potentials are identified as the
soliton solutions of the KdV hierarchy. An $N$-soliton potential has the time $t$ 
and $2N$ positive parameters, $k_1<\cdots<k_N$ and $\{c_j\}$, $j=1,\ldots,N$,
corresponding to $N$ discrete eigenvalues $\{-k_j^2\}$. The eigenfunctions are
elementary functions expressed by the ratio of determinants.
 The Darboux-Crum-Krein-Adler transformation or the Abraham-Moses transformations based
 on eigenfunctions deletions produce lower soliton number potentials with
 modified parameters $\{c'_j\}$. 
 We explore various identities satisfied by the eigenfunctions of the soliton potentials, which reflect the
 uniqueness theorem of  Gel'fand-Levitan-Marchenko equations for separable (degenerate) kernels.
\end{abstract}

\section{Introduction}
\label{intro}

In recent years, exactly solvable potentials in one dimensional quantum mechanics
have aroused resurgent interest thanks to the discovery of certain new solvable
potentials having the exceptional and the multi-indexed 
orthogonal polynomials as the main part of the eigenfunctions.
These new solvable potentials are obtained by rational deformations of known 
solvable potentials;
these are the radial oscillator, P\"oschl-Teller and  Coulomb potentials having 
infinitely many discrete eigenstates and 
Morse, Rosen-Morse, Eckart, hyperbolic P\"oschl-Teller,  
$\sech^2x$ and the hyperbolic symmetric top potentials with finitely many discrete eigenstates.
One conspicuous absence is the reflectionless potentials \cite{KM}, or with
the explicit time dependence the so-called $N$-soliton solutions 
\cite{hirota} of the KdV hierarchy.
For the reflectionless potentials derived by Kay and Moses \cite{KM} in 1956,
all the eigenfunctions are exactly calculable for any number of arbitrarily given eigenvalues
$\{-k_j^2\}$, $j=1,\ldots,N$.
However, their exact solvability does not seem to be widely known among the
present day researchers of the subject.
This is partly because the reflectionless potentials are  {\em not shape invariant\/}
and do not satisfy the well established sufficient condition of exact solvability.

In this paper we address the problem of possible solvable deformations of the
reflectionless potentials in terms of the eigenfunctions, \`a la Darboux-Crum-Krein-Adler
\cite{darb}-\cite{adler}
and Abraham-Moses \cite{A-M,os31}. 
They are known to generate reflectionless potentials \cite{hls}.
Contrary to the naive expectation, these deformations of the general soliton solutions
do not produce a new type of reflectionless potentials, or
new species of soliton solutions.
This is in good  contrast to the known deformation examples like the
multi-index \cite{os25, gomez3} and the exceptional orthogonal polynomials \cite{gomez}--\cite{os28}
cases.
This fact is  consistent with the uniqueness of the reflectionless
potentials as the solutions of Gel'fand-Levitan-Marchenko equations \cite{inv} 
for separable (or degenerate) kernels. (See Appendix of Kay and Moses \cite{KM}.)
The non-deformation, in turn, could be understood as the consequences of 
 many interesting  identities satisfied by the reflections potentials and their eigenfunctions.
We explore these identities as the characteristic properties of the reflectionless
potentials.
An attempt to deform $\sech^2x$ potentials with special $t$ dependence to create
 integer speed solitons was reported recently \cite{hl}.

The present paper is organised as follows.
In section two, the explicit formulas of the reflectionless potential and
the eigenfunctions are
recapitulated for introducing necessary notation and for self-containedness.
An alternative and intuitive derivation of the reflectionless potential and eigenfunctions
is presented.
The relationship between the reflectionless potentials and the soliton solutions
of the KdV hierarchy is explained in some detail.
Section three is the main body of the paper. 
Deformations of reflectionless potential by deleting single and multiple eigenstates via
Darboux-Crum-Krein-Adler and Abraham-Moses transformations are performed explicitly.
Several interesting Wronskian identities among the eigenfunctions of reflectionless (soliton)
potentials are derived instead of new species of reflectionless potentials.
The final section is for a summary and comments. The basic formulas of multiple Darboux and 
Abraham-Moses transformations are recapitulated in \S\ref{sec:darb} 
and \S\ref{sec:A-M}, respectively, for reference purposes.

\section{Reflectionless  potential and its eigenfunctions}
\label{sec:eigen}

Here we recapitulate the essence of the reflectionless potential and the corresponding eigenfunctions.
Since most of the results are well-known for more than forty years, 
we will not give the details of the
derivation and refer to the original paper and related references \cite{KM,hirota,td}.
Let us start with a reflectionless potential $U_N(x)$ in one dimensional quantum mechanics.
It is defined on the entire real line $-\infty<x<\infty$ and it vanishes at $\pm\infty$. 
Its scattering wave solution is reflectionless:
\begin{align}
  &\mathcal{H}=-\frac{d^2}{dx^2}+U_N(x),
  \label{schr}\\
 & \mathcal{H}\psi_k(x)=k^2 \psi_k(x),
 \quad 
 \psi_k(x)\sim \left\{
\begin{array}{rc}
t(k)e^{ikx}   & x\to +\infty\\
e^{ikx}   & x\to -\infty  
\end{array},
\right.
\quad k>0.
\label{scat}
\end{align}
It has $N$ arbitrarily given discrete eigenvalues, 
\begin{align}
  &\mathcal{H}\phi_{N,j}(x)=\mathcal{E}_j\phi_{N,j}(x),\quad \mathcal{E}_j=-k_j^2,
  \quad j=1,\ldots,N,
    \label{sheq}
\end{align}
    with $0<k_1<k_2<\cdots<k_N$, corresponding to the the poles of 
    the transmission amplitude $t(k)$, 
    on the positive imaginary $k$-axis,  $k=ik_j$, $j=1,\ldots,N$.
According to Kay and Moses \cite{KM}, the potential is {\em everywhere negative}, $U_N(x)<0$,  
and it has an expression 
\begin{align}
 U_N(x)&\eqdef-2\partial_x^2\log u_N(x),
 \label{logot}\\
u_N(x)&\eqdef\det A_N(x),\quad
(A_N(x))_{m\,n}\eqdef\delta_{m\,n}+\frac{c_m e^{-(k_m+k_n)x}}{k_m+k_n},
\quad m,n=1,\ldots,N,
\label{logot2}
\end{align}
in which $\{c_m\}$ are arbitrary positive parameters.
For real $x$ and positive $\{k_j\}$ and $\{c_j\}$, the $N\times N$ matrix 
$A_N(x)$ is positive definite.
The {\em logarithmic potential} $u_N(x)$ has a simple expansion \cite{hirota} 
\begin{align}
u_N(x)&=\sum_\mu\exp\left[\sum_{j=1}^N \mu_j\eta_j+\sum_{j<l}a_{j\,l}\mu_j\mu_l\right],
\label{uNform}\\
 e^{\eta_j}&\eqdef \frac{c_j}{2k_j}e^{-2k_jx},
 \quad e^{a_{j\,l}}\eqdef \frac{(k_j-k_l)^2}{(k_j+k_l)^2},\quad 
\quad \mu_j=0,1,
\label{phaseshift}
\end{align}
in which $\sum_\mu$ means a summation over $2^{N}$  terms of 
$\mu_1=0,1$, $\mu_2=0,1$,\ldots, $\mu_N=0,1$.
This expression also shows that $u_N(x)$ is  positive for real $x$ and positive 
$\{k_j\}$ and $\{c_j\}$.

\medskip
For {\em consecutive positive integers\/} $\{k_j\}$, 
and special values of the parameters $\{c_j\}$,
\begin{align}
k_j&=j, \quad c_j=\frac{(N+j)!}{j!(j-1)!(N-j)!},\quad j=1,\ldots,N,\\
u_N(x)&=e^{-N(N+1)x}(1+e^{2x})^{N(N+1)/2},\quad U_N(x)=-N(N+1)\sech^2x,
\end{align}
the general reflectionless potential $U_N(x)$ \eqref{logot} reduces to the reflectionless 
$\sech^2x$ potential,
which is known to be exactly solvable with the eigenvalues 
$\mathcal{E}_j=-j^2$, $j=1,\ldots,N$.

Except for possible complex zeros of $u_N(x)$, the potential $U_N(x)$ is {\em holomorphic\/}.
At a complex simple zero $x_0$ of of $u_N(x)$,  $u_N(x)=(x-x_0)r_N(x)$, $r_N(x_0)\neq0$,
the reflectionless potential $U_N(x)$ \eqref{logot} has a {\em regular singularity\/}
\begin{equation}
U_N(x)=\frac{2}{(x-x_0)^2}+O(x-x_0),
\end{equation}
with the {\em characteristic exponents\/} $2$ and $-1$. 
This means  that the solutions of the Schr\"odinger equations with the potential 
$U_N(x)$ are generically
{\em monodromy free\/} on the complex $x$-plane.

The special form of the reflectionless potential \eqref{logot} means that  
$U_N(x)$  can be derived
by {\em multiple Darboux transformations\/} from the {\em trivial potential\/} $U\equiv0$.
The Schr\"odinger equation with $U\equiv0$ has {\em square non-integrable solutions\/}
\begin{align}
&\psi_j(x)\eqdef e^{k_jx}+{\tilde{c}_j}e^{-k_jx},
\quad 0<k_1<k_2<\cdots<k_N,\quad (-1)^{j-1}\tilde{c}_j>0,
\label{cjsign}\\
&-\partial_x^2\psi_j(x)=-k_j^2\psi_j(x),\quad j=1,\ldots,N.
\end{align}
Their inverses $\{1/\psi_j(x)\}$ are locally square integrable at $x=\pm\infty$.
With the above sign of the parameters $\{\tilde{c}_j\}$ \eqref{cjsign}  the Wronskian 
($\text{W}[f_1,\cdots,f_n](x)\eqdef\det\bigl(\partial_x^{j-1}
f_k(x)\bigr)_{1\le j,k\le n}$) of these seed solutions $\{\psi_j\}$ is positive and it gives $u_N(x)$ 
upto a factor which is annihilated by $\partial_x^2$
after taking the logarithm:
\begin{align}
&\text{W}[\psi_1,\cdots,\psi_N](x)=\prod_{j>l}^N(k_j-k_l)\cdot e^{\sum_{j=1}^Nk_jx}u_N(x),\\
&U_N(x)=-2\partial_x^2\log\text{W}[\psi_1,\cdots,\psi_N](x)=-2\partial_x^2\log u_N(x).
\label{darbN}
\end{align}
Here we have redefined the coefficient of $e^{-2k_jx}$ in $u_N(x)$ to be $c_j/(2k_j)$, $c_j>0$.
Similar derivation of the reflectionless potential, without the eigenfunctions,  was reported
more than twenty years ago \cite{matv-sal}.
The general theory of Darboux transformation says that the above constructed $U_N(x)$ \eqref{darbN} 
has $N$-discrete eigenvalues $\mathcal{E}_j=-k_j^2$ with the eigenfunctions
\begin{equation}
\phi_{N,j}(x)\propto \frac{\text{W}[\psi_1,\cdots,\breve{\psi}_j,\cdots,\psi_N](x)}{\text{W}[\psi_1,\cdots,\psi_N](x)},\quad 
j=1,\ldots,N,
\end{equation}
in which $\breve{\psi}_j$ means that $\psi_j(x)$ is excluded from the Wronskian. 
For derivation, see \eqref{varphiM}.
By the same multiple Darboux transformation, the plane wave solution 
$e^{ikx}$ ($k>0$) of the $U\equiv0$
Schr\"odinger equation is mapped to
\begin{align}
e^{ikx}\to \frac{\text{W}[\psi_1,\cdots,\psi_N,e^{ikx}](x)}{\text{W}[\psi_1,\cdots,\psi_N](x)}
\sim\left\{
\begin{array}{ccc}
\prod_{j=1}^N(ik-k_j)\cdot e^{ikx}  &  x\to +\infty\\[4pt]
\prod_{j=1}^N(ik+k_j)\cdot e^{ikx}    &   x\to -\infty
\end{array}
\right.,
\end{align}
as the Wronskian of  exponential functions is a van der Monde determinant:
\begin{equation*}
\text{W}[e^{\alpha_1x},e^{\alpha_2x},\ldots,e^{\alpha_M x}](x)=
\prod_{1\le k<j\le M}(\alpha_j-\alpha_k)\cdot e^{\sum_{j=1}^M \alpha_j x}.
\end{equation*}
This scattering wave solution   has reflectionless asymptotic behaviour, 
which is consistent with \eqref{scat}.
This is an alternative  derivation of the reflection potential \eqref{logot}.  
{\em Its reflectionless property  and the exact solvability are quite  intuitively understood\/}.

\bigskip
Now we will comment on the relationship between the reflectionless potential \eqref{logot},\eqref{logot2} and the $N$-soliton
solution of the KdV hierarchy.
By construction, the eigenvalues $\{-k_j^2\}$ are independent of the parameters $\{c_j\}$.
Any continuous change of $\{c_j\}$ generate continuous {\em iso-spectral deformation\/} of the
reflectionless potential $U_N(x)$. A special choice of $t$-dependence
\begin{equation}
c_j\to c_je^{8k_j^3t},\quad j=1,\ldots,N,
\label{cjtime}
\end{equation}
changes the reflectionless potential $U_N(x)$ to an $N$-{\em soliton solution\/} $U_N(x;t)$ 
of the KdV equation \cite{hirota,td}:
\begin{align}
 U_N(x;t)&\eqdef-2\partial_x^2\log u_N(x;t),
 \label{Nsoli}\\
u_N(x;t)&\eqdef\det A_N(x;t),\quad
(A_N(x;t))_{m\,n}\eqdef\delta_{m\,n}+\frac{c_m e^{-(k_m+k_n)x+8k_m^3t}}{k_m+k_n},
\label{Nsoli1}\\ 
%
0&=\partial_tU_N-6U_N\partial_xU_N+\partial_x^3U_N.
\label{kdveq}
\end{align}
From now on we will abuse the language and call both $U_N(x)$ \eqref{logot}, 
\eqref{logot2} and $U_N(x;t)$ \eqref{Nsoli}, \eqref{Nsoli1} $N$-{\em soliton solutions\/}. 
In the KdV equation \eqref{kdveq}, the $x$ and $t$ dependence of 
$U_N(x;t)$ is suppressed for simplicity of presentation.
More general time dependence
\begin{equation}
c_j\to c_j\exp[\sum_{n=1}^\infty(2k_j)^{2n+1}t_{2n+1}],\quad j=1,\ldots,N,
\label{cjtime1}
\end{equation}
generates the $N$-{\em soliton solution\/}  of the KdV {\em hierarchy\/}.
Here $t_{2n+1}$ ($t_3\equiv t$) is the time parameter corresponding to the 
$n$-th involutive Hamiltonian of the KdV hierarchy.
As the solution of the non-linear KdV equation \eqref{kdveq}, the overall normalisation
of $U_N(x)$
including the sign is immaterial, since it can be absorbed by the rescaling of the coefficient
of the nonlinear term.
It should be stressed, however, that the overall scale with the  sign $-2$ is essential 
for the potential of the Schr\"odinger equation as shown  above. 

The eigenfunctions $\{\phi_{N,j}(x)\}$ of the reflectionless potential $U_N(x)$ \eqref{logot}, \eqref{logot2}
have a simple expression as the ratio of determinants:
\begin{align}
\phi_{N,j}(x)&\eqdef\frac{\tilde{u}_{N,j}(x)}{u_N(x)}e^{-k_jx},
\quad  \tilde{u}_{N,j}(x)\eqdef \det \widetilde{A}_{N,j}(x),
 \quad  j=1,\ldots,N,
    \label{phinj}\\
( \widetilde{A}_{N,j}(x))_{m\,n}&\eqdef
\delta_{m\,n}+\frac{k_j-k_m}{k_j+k_m}\frac{c_m e^{-(k_m+k_n)x}}{k_m+k_n},
\qquad \qquad m,n=1,\ldots,N.
\label{Atil}
\end{align}
In other words, $\tilde{u}_{N,j}(x)$ is obtained from $u_N(x)$ by the replacement 
\begin{equation}
\tilde{u}_{N,j}(x):\,c_m\to c_m\times\frac{k_j-k_m}{k_j+k_m}, \quad m=1,\ldots,N.
\label{repl}
\end{equation}
It is easy to see that the eigenfunction $\phi_{N,j}(x)$ is 
square integrable with the proper asymptotic behaviour:
\begin{equation}
\phi_{N,j}(x)\sim \left\{
\begin{array}{rc}
e^{-k_jx}   & x\to +\infty\\
\text{constant}\times e^{+k_jx}   & x\to -\infty  
\end{array}.
\right.
\end{equation}
The {\em groundstate\/} eigenfunction $\phi_{N,N}(x)$ 
corresponding to the lowest eigenvalue $-k_N^2$ is
{\em positive\/} $\phi_{N,N}(x)>0$ since the matrix $\widetilde{A}_{N,N}(x)$ is positive definite. 
It needs no explanation that the eigenfunctions for the time-dependent potential 
$U(x;t)$ \eqref{Nsoli}, 
{\em i.e.\/} the $N$-soliton solution, are obtained from \eqref{phinj}, \eqref{Atil} 
by the same replacement 
\eqref{cjtime} or \eqref{cjtime1}.

\section{Deformations and identities}
\label{deform}

Here we will discuss possible deformations of the reflectionless potentials 
and soliton solutions, 
the main theme of the present paper.
By using the eigenfunctions, one can construct an $(N-M)$-soliton solution from 
an $N$-soliton solution.
One naively expects that the resulting $(N-M)$-soliton would retain the dependence on the 
original $2N$ parameters, creating new species of solitons.
However, several attempts in terms of Darboux and Abraham-Moses transformations 
have failed to produce such new types of solitons.
Non-existence of  new species of solitons is consistent with the uniqueness theorem
of reflectionless potentials (see, the Appendix of Kay-Moses original paper \cite{KM}).
On the other hand, it means various identities satisfied by the soliton solutions, 
which do not seem to be widely recognised or discussed. 
The eigenfunctions of exactly solvable quantum mechanical systems are known to satisfy
various interesting identities.
See \cite{os29} for the Wronskian  identities satisfied by 
the Hermite, Laguerre and Jacobi polynomials.
Similar Casoratian identities satisfied by the classical orthogonal polynomials 
obeying second order difference equations, those for the Wilson and 
Askey-Wilson polynomials, are reported in \cite{os30}.

Let us start with the standard Darboux-Crum \cite{crum} transformation by using the
ground state eigenfunction as the seed solution (see \eqref{1pot}):
\begin{equation}
U_N(x)\to U_N^{(1)}(x)=U_N(x)-2\partial_x^2\log\phi_{N,N}(x)=
-2\partial_x^2\log\tilde{u}_{N,N}(x).
\end{equation}
The resulting reflectionless potential is an $N-1$ soliton solution
depending on $2(N-1)$ parameters $\{k_m,c^{(1)}_m\}$, $m=1,\ldots,N-1$ obtained 
from those of the original one by replacement
\begin{equation}
c_m\to c_m^{(1)}\eqdef c_m\times\frac{k_N-k_m}{k_N+k_m}, \qquad m=1,\ldots,N-1.
\label{replN}
\end{equation}
This replacement rule could be interpreted as generalized shape invariance.

Next we deform the $N$ soliton solution by using $M$ distinct eigenfunctions specified by $\mathcal{D}\eqdef\{d_1,\ldots,d_M\}\subset\{1,\ldots,N\}$,  that is by using
\begin{equation}
\{\phi_{N,d_1}(x),\ldots, \phi_{N,d_M}(x)\}.
\label{eigenseed}
\end{equation}
In other words, the solitons having the  parameters $\{k_{d_1},\ldots,k_{d_M}\}$  
are deleted from the original $N$-soliton solution with the  parameter set
$\{k_1,\ldots,k_N\}$:
\begin{equation*}
\{k_{d_1},\ldots,k_{d_M}\}\subset \{k_1,\ldots,k_N\}.
\end{equation*}
This type of deformation is called Krein-Adler transformation \cite{adler}.
The result is the $N-M$ soliton solution with the set of parameters 
$\{k_1,\ldots,k_N\}\backslash\{k_{d_1},\ldots,k_{d_M}\}$ and $\{c_m^{(M)}\}$:
\begin{align}
U_N^{(M)}(x)&=-2\partial_x^2\log\tilde{u}_{N,\mathcal{D}}(x),
\label{kapot}\\
\tilde{u}_{N,\mathcal{D}}(x):\,c_m^{(M)}&
\eqdef c_m\times\prod_{j=1}^M\frac{k_{d_j}-k_m}{k_{d_j}+k_m},
\label{cmM}
\end{align}
in which $\tilde{u}_{N,\mathcal{D}}(x)$ is obtained from $u_N(x)$ \eqref{logot2} 
by replacing $c_m$ with the above $c_m^{(M)}$ \eqref{cmM}.
This means the following Wronskian identity:
\begin{align}
&\text{W}[\phi_{N,d_1},\ldots, \phi_{N,d_M}](x)\propto \frac{\tilde{u}_{N,\mathcal{D}}(x)e^{-\sum_{j=1}^Mk_{d_j}x}}{u_N(x)},
\label{wronid}
\end{align}
as \eqref{Mpot} says
\begin{align*}
 U_N^{(M)}(x)=U_N(x)-2\partial_x^2\log\text{W}[\phi_{N,d_1},\ldots, \phi_{N,d_M}](x)=-2\partial_x^2\log\tilde{u}_{N,\mathcal{D}}(x).
\end{align*}
The positivity of $\tilde{u}_{N,\mathcal{D}}(x)$ is guaranteed if $\mathcal{D}$ 
is chosen to satisfy the conditions \cite{adler}:
\begin{equation}
\prod_{j=1}^M({d}_j-m)\ge0,\quad m=1,\ldots,N.
\end{equation}
With these conditions it is trivial to see the non-negativeness of 
 $c_m^{(M)}\ge0$  \eqref{cmM} for positive $c_m>0$.
These conditions are easily satisfied if $\mathcal{D}$ consists of 
one or many pairs of two consecutive integers.

\bigskip
Next we create an $N-M$-soliton solution from the $N$-soliton solution 
\eqref{logot},\eqref{logot2} by using eigenstate deleting Abraham-Moses transformations 
\eqref{UMAM}--\eqref{Fdef}.
In this case the parameters $\{e_j\}$ are not arbitrary but are the norm of the seed functions
$e_j=(\varphi_j,\varphi_j)$, which are the eigenfunctions.
Here we use the standard notation for the inner product, 
$(f,g)\eqdef\int_{-\infty}^\infty\!f(x)g(x)dx$.
This is an artificial constraint to make the formulas \eqref{UMAM}--\eqref{Fdef} look 
symmetrical for addition  and deletion. For $j\neq l$, $(\varphi_j,\varphi_l)=0$ 
because of the orthogonality of the eigenfunctions and we obtain a positive 
definite expression of $\mathcal{F}_M$ \eqref{UMAM}:
\begin{equation}
\left(\mathcal{F}_M\right)_{j\,l}=e_j\delta_{j\,l}-\langle\varphi_j,\varphi_l\rangle
=(\int_{-\infty}^\infty-\int_{-\infty}^x)\varphi_j(y)\varphi_l(y)dy
=\int_x^\infty \!\!\varphi_j(y)\varphi_l(y)dy.
\label{Fint}
\end{equation}
(See \eqref{lang} for the definition of $\langle f,g\rangle$.)
In this form the deformed potential $U^{(M)}(x)$ \eqref{UMAM},\eqref{Fdef} 
is independent of the normalisation of the eigenfunctions.
Let us use $M$ distinct eigenfunctions \eqref{phinj} specified by 
$\mathcal{D}\eqdef\{d_1,\ldots,d_M\}\subset\{1,\ldots,N\}$, that is by using 
$\{\phi_{N,d_1}(x),\ldots, \phi_{N,d_M}(x)\}$ \eqref{eigenseed}.
It is interesting to see that the integrand of $(\mathcal{F}_M)_{j\,l}$ \eqref{Fint} is a 
derivative of a function of the eigenfunction type as
\begin{align}
&\phi_{N,j}^2(x)=-\partial_x\left(\frac{\tilde{w}_{N,j}(x)}{u_N(x)}\cdot\frac{e^{-2k_jx}}{2k_j}\right),
\label{sqform}\\
&\phi_{N,j}(x)\phi_{N,l}(x)=-\partial_x\left(\frac{\tilde{v}_{N;j,l}(x)}{u_N(x)}
\cdot\frac{e^{-(k_j+k_l)x}}{k_j+k_l}\right),\quad \tilde{v}_{N;j,j}(x)\equiv \tilde{w}_{N,j}(x),
\label{biliform}
\end{align}
in which $\tilde{w}_{N,j}(x)$ and $\tilde{v}_{N;j,l}(x)$ are obtained from $u_N(x)$ 
by the following replacements of $c_m$:
\begin{align}
\tilde{w}_{N,j}(x):\, c_m\to  c_m\times\frac{(k_j-k_m)^2}{(k_j+k_m)^2},
\qquad 
\tilde{v}_{N;j,l}(x):\, c_m\to c_m\times \frac{k_j-k_m}{k_j+k_m}\cdot\frac{k_l-k_m}{k_l+k_m}.
\end{align}
The above factor in $\tilde{w}_{N,j}$ is the {\em scattering phase shift} 
of the $j$-th and $m$-th solitons and it is the square of the factor appearing in 
$\tilde{u}_{N,j}$ of the eigenfunction \eqref{phinj}.
The $N-1$-soliton solution obtained by an eigenfunction deleting 
Abraham-Moses transformation
\eqref{u1def} using $\phi_{N,j}(x)$ is
\begin{equation}
U_N(x)\to U_N^{(1)}(x)=U_N(x)-2\partial_x^2\log\int_x^\infty\!\!\phi^2_{N,j}(y)dy=
-2\partial_x^2\log\tilde{w}_{N,j}(x).
\end{equation}
By repeating this process $M$ times in terms of the  eigenfunctions specified by 
$\mathcal{D}$ \eqref{eigenseed} is
\begin{equation}
U_N(x)\to U_N^{(M)}(x)=
-2\partial_x^2\log\tilde{w}_{N,\mathcal{D}}(x), 
\label{MdelAM}
\end{equation}
in which $\tilde{w}_{N,\mathcal{D}}(x)$ is obtained from $u_N(x)$ by the replacement
\begin{equation}
\tilde{w}_{N,\mathcal{D}}(x):\, c_m\to  c_m\times
\prod_{j=1}^M\frac{(k_{d_j}-k_m)^2}{(k_{d_j}+k_m)^2}.
\label{cmMAM}
\end{equation}
This in turn means a determinant identity
\begin{align}
\text{det}\left(\int_x^\infty\!\!\phi_{N,d_j}(y)\phi_{N,d_l}(y)dy\right)_{1\le j,l\le M}
&=\text{det}\left(\frac{\tilde{v}_{N;d_j,d_l}(x)}{u_N(x)}\cdot
\frac{e^{-(k_{d_j}+k_{d_l})x}}{k_{d_j}+k_{d_l}}\right)_{1\le j,l\le M}
\label{detid}\\
&\propto 
\frac{\tilde{w}_{N,\mathcal{D}}(x)}{u_N(x)} e^{-2\sum_{j=1}^M k_{d_j}x}.
\label{detid1}
\end{align}

\bigskip
We now turn to the clarification of the role played by the eigenstate adding
Abraham-Moses transformations on the $N$-soliton solution \eqref{logot},\eqref{logot2}.
As remarked in \S\ref{sec:A-M}, these transformations are exactly iso-spectral when
the seed solutions are the eigenfunction themselves. In order to fix the interpretation of the
parameters $\{e_j\}$ in \eqref{u1def}--\eqref{Fdef},  let us use the {\em normalised\/} 
seed solutions $\hat{\varphi}_j$, $(\hat{\varphi}_j,\hat{\varphi}_j)=1$.
For the $N$-soliton solution, they are
\begin{equation}
\hat{\phi}_{N,j}(x)\eqdef \sqrt{c_j}\phi_{N,j}(x),\quad
(\hat{\phi}_{N,j},\hat{\phi}_{N,j})=1,\quad j=1,\ldots,N,
\end{equation}
for the  relation \eqref{sqform} gives a simple way to normalise the eigenfunctions:
\begin{align}
&\frac{\tilde{w}_{N,j}(x)}{u_N(x)}\cdot\frac{e^{-2k_jx}}{2k_j}
\to \left\{
\begin{array}{cl}
 0 &   x\to+\infty   \\[2pt]
  \frac1{c_j}&   x\to-\infty 
\end{array}
\right.,\quad (\phi_{N,j},\phi_{N,j})=\frac1{c_j}.
\end{align}
The one eigenstate addition by using $\hat{\phi}_{N,j}(x)$ goes as follows \eqref{u1def}:
\begin{align}
e_j+\langle\hat{\phi}_{N,j},\hat{\phi}_{N,j}\rangle
&=e_j+1-\frac{\tilde{w}_{N,j}(x)}{u_N(x)}\cdot\frac{c_je^{-2k_jx}}{2k_j}\n
&=(e_j+1)\frac{\left\{n_{N,j}(x)+\frac{e_j}{e_j+1}\frac{c_j}{2k_j}e^{-2k_jx}\tilde{w}_{N,j}(x)\right\}}{u_N(x)}\n
&=(e_j+1)\frac{\tilde{z}_{N,j}(x)}{u_N(x)},
\end{align}
in which $\tilde{z}_{N,j}(x)$ is obtained from $u_N(x)$ by the replacement:
\begin{equation}
\tilde{z}_{N,j}(x):\,c_j\to \frac{e_j}{e_j+1}c_j.
\end{equation}
In short, the eigenstate adding Abraham-Moses transformation in terms of the $j$-th 
eigenfunction $\hat{\phi}_{N,j}(x)$ and $e_j$  does not introduce a new independent parameter.
 It simply rescales the corresponding $j$-th parameter $c_j$ 
to $\frac{e_j}{e_j+1}c_j$, namely $u_N(x)$ to $\tilde{z}_{N,j}(x)$.
Here we have used a simple identity of the logarithmic potential $u_N(x)$:
\begin{equation}
u_N(x)=u_{N,j}(x)+\frac{c_j}{2k_j}e^{-2k_jx}\tilde{w}_{N,j}(x),
\quad u_{N,j}(x)\eqdef \left.u_N(x)\right|_{c_j\to0},
\end{equation}
which is obvious from the expansion formula \eqref{uNform}--\eqref{phaseshift}.

By repeating this process $M$ times in terms of the  eigenfunctions specified by $\mathcal{D}$ 
\eqref{eigenseed}, we obtain
\begin{equation}
U_N(x)\to U_N^{(M)}(x)=
-2\partial_x^2\log\tilde{z}_{N,\mathcal{D}}(x), 
\label{MaddAM}
\end{equation}
in which $\tilde{z}_{N,\mathcal{D}}(x)$ is obtained from $u_N(x)$ by the replacement
\begin{equation}
\tilde{z}_{N,\mathcal{D}}:\, c_{d_j}\to  \frac{e_{d_j}}{e_{d_j}+1}c_{d_j},\quad j=1,\ldots,M,
\quad c_l\to c_l,\quad l\notin\mathcal{D}.
\label{cmaddAM}
\end{equation}
This in turn means a determinant identity
\begin{align}
&\text{det}\left(e_{d_j}\delta_{j\,l}+\langle\hat{\phi}_{N,d_j},\hat{\phi}_{N,d_j}\rangle\right)_{1\le j,l\le M}\n
&=\text{det}\left((e_{d_j}+1)\delta_{j\,l}-\frac{\tilde{v}_{N;d_j,d_l}(x)}{u_N(x)}\cdot
\frac{\sqrt{c_{d_j}c_{d_l}}e^{-(k_{d_j}+k_{d_l})x}}{k_{d_j}+k_{d_l}}\right)_{1\le j,l\le M}
\label{adddetid}\\
&\propto 
\frac{\tilde{z}_{N,\mathcal{D}}(x)}{u_N(x)} .
\label{adddetid1}
\end{align}
This type of iso-spectral transformations were reviewed in \S7 of \cite{susyqm},
with the integer solitons of the $\sech^2x$ potential, special cases of the general soliton solutions.

As is known \cite{schn-leeb,sam}  the one eigenstate ($\varphi$) adding/deleting 
Abraham-Moses transformation is obtained by the ordinary Darboux transformation
in terms of  $\varphi$ followed by another in terms of 
$\bar{\varphi}^{(1)}\eqdef{\varphi}^{-1}(e\pm\langle\varphi,\varphi\rangle)$, 
which is a particular solution of the first deformed Hamiltonian with 
$U^{(1)}=U-2\partial_x\log|\varphi|$.
Thus it is also called a {\em binary Darboux transformation\/} in some research group.

\section{Summary and comments}
\label{summary}

Contrary to the naive expectation, the deformation of the $N$-soliton solution
in terms of $M$ distinct eigenfunctions specified by $\mathcal{D}=\{d_1,\ldots,d_M\}$
does not produce new species of $(N-M)$-soliton solutions depending on $2N$ independent
parameters.
The obtained $(N-M)$-soliton solution depends on $2(N-M)$ independent parameters:
\begin{align}
\{k_1,\ldots,k_N\}\backslash\{k_{d_1},\ldots,k_{d_M}\},\quad
c^{(M)}_m=c_m\times \prod_{j=1}^M\left(\frac{k_{d_j}-k_m}{k_{d_j}+k_m}\right)^\Xi,
\quad m\in\{1,\ldots,N\}\backslash\mathcal{D},
\end{align}
in which $\Xi=1$ for the Krein-Adler (multiple Darboux) 
transformation \eqref{kapot}--\eqref{cmM}
and $\Xi=2$ for the multiple eigenstate deleting Abraham-Moses transformation 
\eqref{MdelAM}--\eqref{cmMAM}.
The failure to generate new species of soliton solutions, in turn, implies various Wronskian (determinant) identities \eqref{wronid}, \eqref{detid}--\eqref{detid1}, \eqref{adddetid}--\eqref{adddetid1} in a similar way as 
other exactly solvable quantum mechanical systems \cite{os29, os30}.
We did not address the problem of solution generating transformations
of non-linear PDE's, {\em e.g.\/} B\"acklund transformations.

After the discovery of solitons of KdV equation \cite{kruskal,lax},
more general scheme of inverse scattering theory were developed by AKNS-ZS \cite{akns,zs},
which covered the modified KdV, sine-Gordon and non-linear Schr\"odingier equations 
among others.
It would be interesting to pursue similar goals with these soliton solutions \cite{hirota1};
if  they are exactly solvable, if their deformations generate new types of soliton solutions
or result in various identities, the relationship with infinitely many conserved quantities
and the corresponding Hamiltonian structures, geometrical interpretations, etc \cite{wadati, pssurf}.

Another challenge is the discretized solitons, which pose almost the same questions 
as above from the point of view of discrete quantum mechanics \cite{os24}.
The analogues of Darboux-Crum-Krein-Adker transformations  in discrete quantum mechanics 
are known \cite{gos} and the analogues of the Wronskian identities, 
Casoratian identities, for exactly solvable systems are also reported \cite{os30}.

Most of the shape invariant and exactly solvable quantum mechanical systems
can also be solved exactly in the {\em Heisenberg picture} \cite{os7}.
It is interesting to try and find  the Heisenberg operator solutions for the reflectionless potentials.

\section*{Acknowledgements}
R.\,S. thanks Jen-Chi Lee and Choon-Lin Ho for useful discussion and for the hospitality at
National Chiao-Tung University, National Center for Theoretical Sciences (North) 
and  National Taiwan University.
He also thanks K.\, Takasaki, S.\,Tsujimoto and E.\,Date for useful discussion.
 R.\,S. is supported in part by Grant-in-Aid for Scientific Research
from the Ministry of Education, Culture, Sports, Science and Technology
(MEXT),  No.22540186.


\bigskip

\bigskip

\appendix{\Huge{\textbf{{Appendix }}}}

%
%
%
\section{Deformation schemes}
\label{deformscheme}
Here we provide  a brief summary of Darboux transformations and other methods of 
deformation of the potentials and solutions of Schr\"odinger equations;
\begin{align}
  \mathcal{H}=-\frac{d^2}{dx^2}+U(x),\quad
  \mathcal{H}\psi(x)&=\mathcal{E}\psi(x)\ \quad
  \bigl(\mathcal{E},U(x)\in\mathbb{C}\bigr),
  \label{schr2}\\
    \mathcal{H}\varphi_j(x)&=\tilde{\mathcal{E}}_j\varphi_j(x)\quad
  (\tilde{\mathcal{E}}_j\in\mathbb{C}\ ;\ j=1,2,\ldots,M),
  \label{scheq2}
\end{align}
The functions $\{\varphi_j(x), \tilde{\mathcal{E}}_j\}$ ($j=1,2,\ldots,M$) 
are called {\em seed} solutions.
The subsequent two subsections are for self-containedness.

%
%
\subsection{Multiple Darboux transformation}
\label{sec:darb}
By picking up one of the above seed solutions, say  $\varphi_1(x)$,
we form new functions with the above solution $\psi(x)$ and the rest
of $\{\varphi_l(x),\tilde{\mathcal{E}}_l\}$ ($l\neq 1$):
\begin{equation}
  \psi^{(1)}(x)\eqdef\frac{\text{W}[\varphi_1,\psi](x)}{\varphi_1(x)}
  =\frac{\varphi_1(x)\partial_x\psi(x)
  -\partial_x\varphi_1(x)\psi(x)}{\varphi_1(x)},\quad
  \varphi^{(1)}_{1,l}(x)\eqdef
  \frac{\text{W}[\varphi_1,\varphi_l](x)}{\varphi_1(x)}.
\end{equation}
It is elementary to show that $\psi^{(1)}(x)$,
$\varphi_1^{-1}(x)\,\bigl(\eqdef\varphi_1(x)^{-1}\bigr)$ and
$\varphi^{(1)}_{1,l}(x)$ are solutions of a new Schr\"odinger equation
of a deformed Hamiltonian $\mathcal{H}^{(1)}$ 
\begin{equation}
  \mathcal{H}^{(1)}=-\frac{d^2}{dx^2}+U^{(1)}(x),\quad
  U^{(1)}(x)\eqdef U(x)-2\partial_x^2\log\bigl|\varphi_1(x)\bigr|,
  \label{1pot}
\end{equation}
with the same energies $\mathcal{E}$, $\tilde{\mathcal{E}}_1$ and
$\tilde{\mathcal{E}}_l$:
\begin{align}
  \mathcal{H}^{(1)}\psi^{(1)}(x)&=\mathcal{E}\psi^{(1)}(x),\quad 
  \mathcal{H}^{(1)}\varphi^{-1}_1(x)=\tilde{\mathcal{E}}_1\varphi^{-1}_1(x),
  \label{newschr}\\
  \mathcal{H}^{(1)}\varphi^{(1)}_{1,l}(x)
  &=\tilde{\mathcal{E}}_l\varphi^{(1)}_{1,l}(x)\ \ (l\neq 1).
  \label{newschr2}
\end{align}
By repeating the above Darboux transformations $M$-times, we obtain new
functions
\begin{align}
  \psi^{(M)}(x)&\eqdef
  \frac{\text{W}[\varphi_1,\varphi_2,\ldots,\varphi_M,\psi](x)}
  {\text{W}[\varphi_1,\varphi_2,\ldots,\varphi_M](x)},
  \label{psiM}\\
  \breve{\varphi}^{(M)}_j(x)&\eqdef
  \frac{\text{W}[\varphi_1,\varphi_2,\ldots,\breve{\varphi}_j,\ldots,
  \varphi_M](x)}
  {\text{W}[\varphi_1,\varphi_2,\ldots,\varphi_M](x)}\ \ (j=1,2,\ldots,M),
  \label{varphiM}
\end{align}
which satisfy an $M$-th deformed Schr\"odinger equation with the energies
$\mathcal{E}$ and $\tilde{\mathcal{E}}_j$ \cite{darb}:
\begin{align}
  &\mathcal{H}^{(M)}=-\frac{d^2}{dx^2}+U^{(M)}(x),\quad
  U^{(M)}(x)\eqdef U(x)-2\partial_x^2\log\bigl|
  \text{W}[\varphi_1,\varphi_2,\ldots,\varphi_M](x)\bigr|,
  \label{Mpot}\\
  &\mathcal{H}^{(M)}\psi^{(M)}(x)=\mathcal{E}\psi^{(M)}(x),\quad
  \mathcal{H}^{(M)}\breve{\varphi}^{(M)}_j(x)
  =\tilde{\mathcal{E}}_j\breve{\varphi}^{(M)}_j(x)\ \ (j=1,2,\ldots,M).
  \label{Mschr}
\end{align}
These multiple Darboux transformations are {\em essentially iso-spectral\/}, 
upto a finite number of added or deleted energy levels depending on the
properties of the seed functions $\{\varphi_j(x), \tilde{\mathcal{E}}_j\}$ ($j=1,2,\ldots,M$).
In order to avoid singularities of the new potential $U^{(M)}(x)$, the Wronskian 
of the seed functions $\text{W}[\varphi_1,\varphi_2,\ldots,\varphi_M](x)$ should not vanish
on the real $x$-axis.

%
%
\subsection{Multiple Abraham-Moses transformation}
\label{sec:A-M}

Next let us introduce the Abraham-Moses transformations \cite{A-M, os31}.
For simplicity of the presentation, we will restrict ourselves to utilise
the solutions and seed solutions which are locally square integrable at $x=-\infty$. 
For a pair of real functions $f$ and $g$,  let us introduce a new function $\langle f,g\rangle$
by integration:
\begin{align}
  \langle f,g\rangle(x)&\eqdef\int_{-\infty}^xdyf(y)g(y)
  =\langle g,f\rangle(x), 
  \label{lang}\\
  \langle f,g\rangle(-\infty)&=0,\quad\langle f,g\rangle(+\infty)
  =(f,g)\eqdef \int_{-\infty}^\infty\! \!f(x)g(x)dx.
\end{align}
Note that $\frac{d}{dx}\langle f,g\rangle(x)=f(x)g(x)$.
Throughout this paper, we use the simplified notation 
$\langle f,g\rangle\equiv \langle f,g\rangle(x)$,
so long as no confusion arises.

For a seed solution, say $\varphi_1$, with the energy $\tilde{\mathcal{E}}_1$,
an Abraham-Moses transformation for adding/deleting one bound state with the energy
$\tilde{\mathcal{E}}_1$, is defined as follows:
\begin{align}
U(x)\to U^{(1)}(x)&\eqdef U(x)-2\partial_x^2
  \log\bigl(e_1\pm \langle\varphi_1,\varphi_1\rangle\bigr),\quad e_1>0,
  \label{u1def}\\
 \varphi_1\to\varphi^{(1)}_1
  &\eqdef\frac{\varphi_1}{e_1\pm\langle\varphi_1,\varphi_1\rangle},
  \qquad \mathcal{H}^{(1)}\varphi^{(1)}_1=\tilde{\mathcal{E}}_1\varphi^{(1)}_1,\\
  \psi\to\psi^{(1)}
  &\eqdef\psi\mp\varphi^{(1)}_1
  \langle\varphi_1,\psi\rangle,
   \quad\ \mathcal{H}^{(1)}\psi^{(1)}=\mathcal{E}\psi^{(1)}.
  \label{psi1defs}
\end{align}
For the eigenstate addition, we choose the upper sign and $e_1>0$ is arbitrary.
In this case a non-square integrable seed solution $(\varphi_1,\varphi_1)=\infty$
is mapped to an eigenstate $\varphi^{(1)}_1$ with the energy $\tilde{\mathcal{E}}_1$.
If $\varphi_1$ is an eigenstate, $\varphi^{(1)}_1$ is also an 
eigenstate with the energy $\tilde{\mathcal{E}}_1$ and its normalsation is changed.
In this case, the transformation is exactly iso-spectral and no eigenstate is added.
For the eigenstate deletion, we choose the lower sign and $e_1$ is the norm of the
eigenstate $\varphi_1$, $e_1\eqdef(\varphi_1,\varphi_1)$.
The transformed state  $\varphi^{(1)}_1$ is no longer square integrable, 
$(\varphi^{(1)}_1, \varphi^{(1)}_1)=\infty$, {\em i.e.\/} the eigenstate is deleted.

By repeating these Abraham-Moses transformations, 
we arrive at multiple eigenstate
adding/deleting transformations \cite{os31,trl}:
\begin{align}
  U^{(M)}(x)&=U(x)-2\partial_x^2
  \log\det\bigl(\mathcal{F}_M\bigr),
  \label{UMAM}\\
  \psi^{(M)}&=\psi\mp\sum_{j,l=1}^M\varphi_j
  \bigl(\mathcal{F}_M^{-1}\bigr)_{j\,l}\langle\varphi_l,\psi\rangle,\quad
   \mathcal{H}^{(M)} \psi^{(M)}=\mathcal{E} \psi^{(M)},
  \label{phiMform}\\
  \varphi^{(M)}_j&=\sum_{l=1}^M
  \bigl(\mathcal{F}^{-1}_M\bigr)_{j\,l}\varphi_l,\quad
   \mathcal{H}^{(M)}  \varphi^{(M)}_j=\tilde{\mathcal E}_j  \varphi^{(M)}_j,
 \quad(j,l=1,\ldots,M),
  \label{varpMform}
\end{align}
in which $\mathcal{F}_M$ is an
$M\times M$ symmetric and positive definite matrix  depending
on the seed solutions $\{\varphi_j\}$ ($j=1,\ldots,M$) defined by:
\begin{equation}
  (\mathcal{F}_M)_{j\,l}\eqdef e_j\delta_{j\,l}
  \pm\langle\varphi_j,\varphi_l\rangle,\quad 
  e_j\left\{
\begin{array}{lc}
 >0\ \text{arbitrary}&   \text{addition}  \\
 \eqdef (\varphi_j,\varphi_j)  &  \text{deletion} 
\end{array}
\right.,
\quad(j,l=1,\ldots,M).
  \label{Fdef}
\end{equation}
The positive definiteness of $\mathcal{F}_M$ guarantees the regularity of the deformed potential.



\end{document}